\documentclass[prc,twocolumn,floatfix,groupedaddress,nofootinbib,showpacs,preprintnumbers,
amsmath,amssymb,amsfonts,superscriptaddress,widetable] {revtex4}
\usepackage{bm}
\usepackage{mathrsfs}
\usepackage{amssymb}
\usepackage{amsmath}
\usepackage{graphicx}
\usepackage{array}
\usepackage{xcolor}
\newcommand{\rhozero}{\rho_{\raisebox{-2.0pt}{\tiny\!0}}}

\newcommand{\msun}{M$_\odot$}
\newcommand{\Lns}{$\Lambda_\star^{1.4}$}
\newcommand{\Rns}{$R_\star^{1.4}$}
\newcommand{\Rskin}{$R_{\rm skin}^{208}$}

\begin{document}
\title{Implications of PREX-II on the equation of state of neutron-rich matter}
\author{Brendan T. Reed}\email{reedbr@iu.edu}
\affiliation{Department of Astronomy, Indiana University, 
                Bloomington, Indiana 47405, USA}
\affiliation{Center for Exploration of Energy and Matter and
              Department of Physics, Indiana University,
              Bloomington, IN 47405, USA}
\author{F. J. Fattoyev}\email{ffattoyev01@manhattan.edu}
\affiliation{Department of Physics, Manhattan College,
                Riverdale, NY 10471, USA}
\author{C. J. Horowitz}\email{horowit@indiana.edu}
\affiliation{Center for Exploration of Energy and Matter and
                  Department of Physics, Indiana University,
                  Bloomington, IN 47405, USA}
\author{J. Piekarewicz}\email{jpiekarewicz@fsu.edu}
\affiliation{Department of Physics, Florida State University,
               Tallahassee, FL 32306, USA}
\date{\today}
\begin{abstract}
 Laboratory experiments sensitive to the equation of state of neutron rich matter
 in the vicinity of nuclear saturation density provide the first rung in a ``density 
 ladder" that connects terrestrial experiments to astronomical observations. In this
 context, the neutron skin thickness of ${}^{208}$Pb (\Rskin) provides a
 stringent laboratory constraint on the density dependence of the symmetry energy. 
 In turn, an improved 
 value of \Rskin\, has been reported recently 
 by the PREX collaboration. Exploiting the strong correlation between \Rskin\, and 
 the slope of the symmetry energy $L$
 within a specific class of relativistic energy density functionals, 
 we report a value of 
 $L\!=\!(106 \pm 37)\,{\rm MeV}$---that systematically overestimates current 
 limits based on both theoretical approaches and experimental measurements. The 
 impact of such a stiff symmetry energy on some critical neutron-star observables is 
 also examined.
\end{abstract}
\smallskip

\maketitle

The updated Lead Radius EXperiment (PREX-II) has delivered on the promise to determine
the neutron radius of ${}^{208}$Pb with a precision of nearly 1\%. By combining the original 
PREX result\,\cite{Abrahamyan:2012gp,Horowitz:2012tj} with the newly announced PREX-II 
measurement, the following value for the neutron skin thickness of ${}^{208}$Pb was  
reported\,\cite{Adhikari:2021phr}:
\begin{equation}
 R_{\rm skin}=R_{n}-R_{p}=(0.283\pm0.071)\,{\rm fm},
 \label{Rskin}
\end {equation} 
where the quoted uncertainty represents a $1\sigma$ error, and $R_{n}$ and $R_{p}$ are the 
root-mean-square radii of the neutron and proton density 
distributions, respectively. Such a purely electroweak measurement is of critical importance in 
constraining both models of nuclear structure as well as the equation of state (EOS) of neutron-rich 
matter in the vicinity of nuclear saturation density ($\rhozero\!\approx\!0.15\,{\rm fm}^{-3}$). In turn, 
the EOS around saturation density provides the first rung in a ``density ladder" that connects 
laboratory experiment to astronomical observations that probe the EOS at higher densities. It is 
the aim of this letter to explore the impact of PREX-II on certain parameters of the EOS that, 
in turn, dictates the behavior of several neutron-star observables.

For two decades the neutron skin thickness of ${}^{208}$Pb has been identified as an ideal 
laboratory observable to constrain the EOS of neutron rich matter, particularly the poorly 
determined density dependence of the symmetry energy\,\cite{Brown:2000,Furnstahl:2001un,
Centelles:2008vu,RocaMaza:2011pm}. The EOS of infinite nuclear matter at zero temperature 
is enshrined in the energy per particle which depends on both the conserved neutron ($\rho_{n}$) 
and proton ($\rho_{p}$) densities; here we assume that the electroweak sector has been ``turned off\,". 
Moreover, it is customary to separate the EOS into two contributions, one that represents the energy 
of  symmetric ($\rho_{n}\!=\!\rho_{p}$) nuclear matter and another one that accounts for the breaking 
of the symmetry. That is,
\begin{equation}
  \frac{E}{A}(\rho,\alpha) -\!M \equiv {\cal E}(\rho,\alpha)
                          = {\cal E}_{\rm SNM}(\rho)
                          + \alpha^{2}{\cal S}(\rho)  
                          + {\cal O}(\alpha^{4}) \,,
 \label{EOS}
\end {equation} 
where $\rho\!=\!(\rho_{n}\!+\!\rho_{p})$ is the total baryon density given by the sum of neutron and
proton densities, and $\alpha\!=(\rho_{n}\!-\!\rho_{p})/\rho$ is the neutron-proton asymmetry. The 
first-order correction to the energy of symmetric nuclear matter ${\cal E}_{\rm SNM}(\rho)$ is 
encoded in the symmetry energy ${\cal S}(\rho) $. The symmetry energy quantifies the increase 
in the energy per particle of infinite nuclear matter for systems with an isospin imbalance (e.g., more 
neutrons than protons). Further, given the preeminent role of nuclear saturation, the energy of 
symmetric nuclear matter and the symmetry energy may be described in terms of a few bulk 
parameters that characterize their behavior around saturation density. In this letter we focus on 
the density dependence of the symmetry energy\,\cite{Piekarewicz:2008nh}:
\begin{equation}
 {\cal S}(\rho) = J + L \frac{(\rho-\rhozero)}{3\rhozero} + \ldots   
\label{EandS}
\end{equation}
The first term ($J$) represents the correction to the binding energy of symmetric nuclear matter,
whereas the second term ($L$) dictates how rapidly the symmetry energy increases with
density. It is the slope of the symmetry energy $L$ that displays a strong correlation to the neutron 
skin thickness of ${}^{208}$Pb. Given that symmetric nuclear matter saturates, namely, its pressure 
vanishes at saturation, the slope of the symmetry energy $L$ is closely related to the pressure of 
pure neutron matter at saturation density. That is,  
\begin{equation}
 P_{\,\rm PNM}(\rhozero)\!\approx\!\frac{1}{3} L\rhozero.  
\label{Ppnm}
\end{equation}

\begin{figure}[ht]
 \centering
 \includegraphics[width=0.5\textwidth]{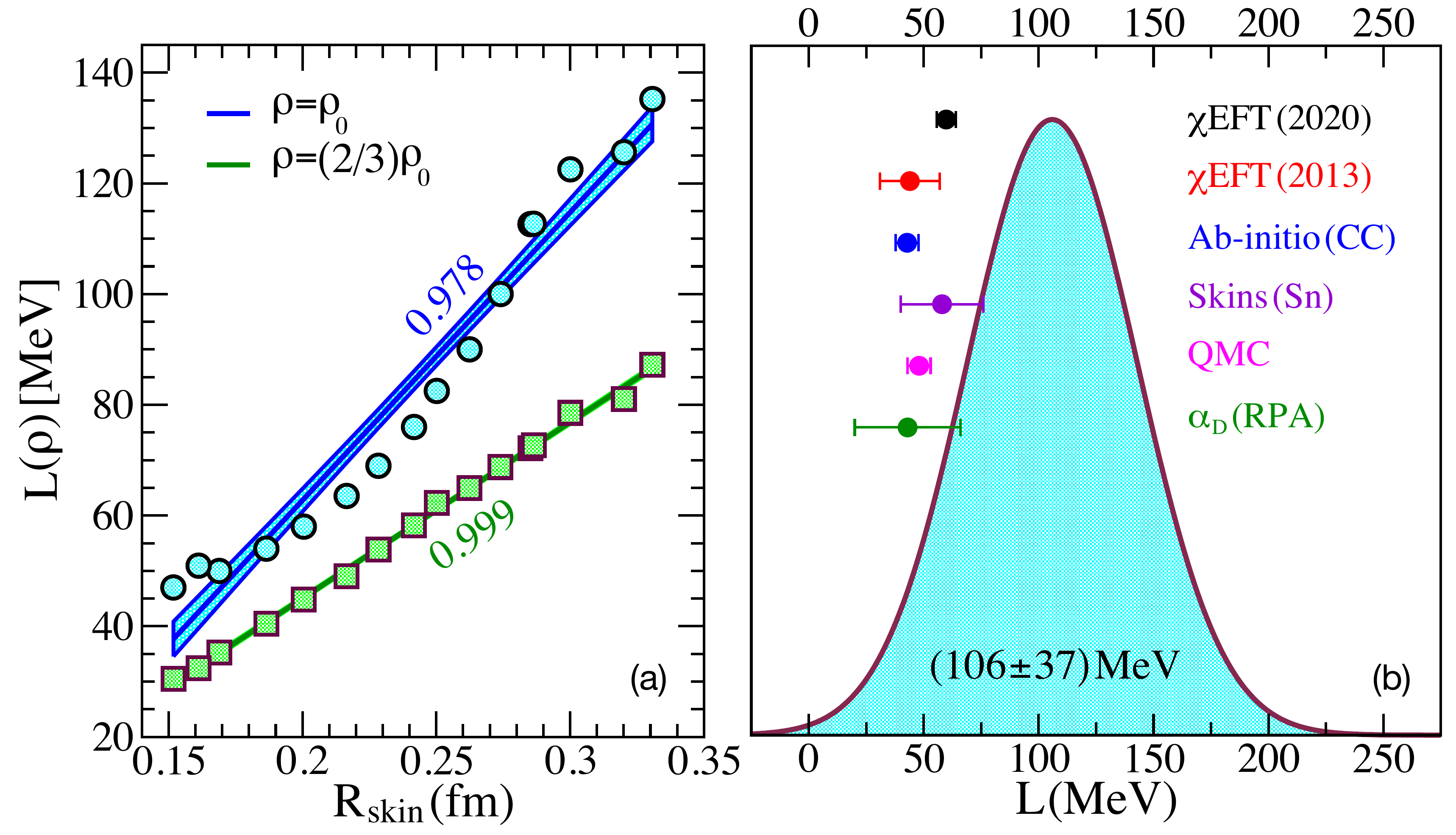}
 \caption{(Color online) \textit{Left}: Slope of the symmetry energy at nuclear saturation density 
 $\rhozero$ (blue upper line) and at $(2/3)\rhozero$ (green lower line) as a function of \Rskin. 
 The numbers next to the lines denote values for the correlation coefficients.
 \textit{Right}: Gaussian probability distribution for the slope of the symmetry energy 
 $L\!=\!L(\rhozero)$ inferred by combining the linear correlation in the left figure with the 
 recently reported PREX-II limit. The six error bars are constraints on $L$ obtained by using 
 different theoretical approaches\,\cite{Zhang:2013wna,Hebeler:2013nza,Drischler:2020hwi,
 Hagen:2015yea,Chen:2010qx,Steiner:2011ft,Gandolfi:2013baa,Roca-Maza:2015eza}.}
\label{Fig1}
\end{figure}

To assess the impact of the combined PREX--PREX-II measurements (henceforth referred simply as 
``PREX-II")---we provide predictions for several observables using a set of 16 covariant energy density 
functionals. These are FSUGold2\,\cite{Chen:2014sca} together with a set of eight systematically varied 
interactions---FSUGold2--L047, L050, L054, L058, L069, L076, L090, L100---with identical isoscalar 
properties as FSUGold2, but isovector properties defined by the associated value of the slope of the 
symmetry energy $L$. For example, FSUGold2=FSUGold2--L113 predicts a slope of the symmetry 
energy of $L\!=\!113\,{\rm MeV}$. Another set of accurately calibrated density functionals is given by
RMF012, RMF016, RMF022, and RMF032\,\cite{Chen:2014mza}, where now the labels are associated 
to the predicted value of \Rskin. For example, RMF032 predicts a neutron skin thickness of 
\Rskin$\!=\!0.32\,{\rm fm}$. Finally, TFa, TFb, and TFc, with \Rskin$\!=\!0.25, 0.30, {\rm and}\, 0.33\,{\rm fm}$, 
respectively, were created to test whether the large central value of \Rskin$\!=\!0.33\,{\rm fm}$ originally 
reported by the PREX collaboration\,\cite{Abrahamyan:2012gp} was incompatible with other laboratory 
experiments and/or astrophysical observations\,\cite{Fattoyev:2013yaa}. We found then, that there was 
no compelling reason to rule out models with large neutron skins. 

From the compilation of all these 16 models one obtains for the binding energy per nucleon 
and the charge radius of ${}^{208}$Pb the following values: $B/A\!=\!7.88\pm0.01\,{\rm MeV}$ 
and $R_{\rm ch}\!=\!5.51\pm0.01\,{\rm fm}$, which compare well against the experimental values
of $B/A\!=\!7.87\,{\rm MeV}$ and $R_{\rm ch}\!=\!5.50\,{\rm fm}$, respectively. A detailed 
description of the fitting protocol---including an explanation of the model and the observables
used in the calibration procedure---may be found in 
Refs.\,\cite{Chen:2014sca,Chen:2014mza,Fattoyev:2013yaa}.
Moreover we
underscore that the models considered here span a wide range of values for both the
neutron skin thickness of ${}^{208}$Pb and the associated slope of the symmetry energy. 
Indeed, the range of adopted values is almost as wide as the one used in the multi-model 
analysis of the sensitivity of the symmetry energy to the electric dipole polarizability and 
weak-charge form factor of both ${}^{48}$Ca and 
${}^{208}$Pb\,\cite{Piekarewicz:2012pp,Reinhard:2013fpa}.


The strong correlation between $L$ and \Rskin in the context of the new PREX-II measurement is illustrated 
in Fig.\,\ref{Fig1}. The left-hand panel displays the well-known correlation between the slope of the symmetry 
energy at saturation density and the neutron-skin thickness of ${}^{208}$Pb. Also shown in Fig.\,\ref{Fig1}(a) 
is the even stronger correlation between $R_{\rm skin}$ and the slope of the symmetry energy at the slightly 
lower density of 
$\widetilde\rhozero\!=\!(2/3)\rhozero\!\approx\!0.1\,{\rm fm}^{-3}$\,\cite{Horowitz:2000xj,Furnstahl:2001un,Ducoin:2010as, Ducoin:2011fy,Horowitz:2014bja, Zhang:2013wna}. At such a lower density, which represents an average value 
between the central and surface densities, the symmetry energy is well constrained by the binding energy of 
heavy nuclei with a significant neutron excess. Relying on the strong $R_{\rm skin}$-$L$ correlation together 
with the improved PREX-II limit, one obtains the gaussian probability distribution for $L$ displayed in Fig.\,\ref{Fig1}(b). 
Using the same analysis on both $J$ and $\widetilde{L}$---the latter representing the slope of the symmetry 
energy at $\widetilde\rhozero$---we derive the following limits:
\begin{subequations}
\begin{align}
 & J = (38.1 \pm 4.7) {\rm MeV}, \\
 & L = (106 \pm 37) {\rm MeV},  \\
 & \widetilde{L} = (71.5 \pm 22.6) {\rm MeV}.
\end{align} 
\label{JandL}
\end{subequations}
As indicated in Fig.\,\ref{Fig1}(b), these limits are systematically larger than those obtained using 
either purely theoretical approaches or extracted from a theoretical interpretation of experimental 
data\,\cite{Zhang:2013wna,Hebeler:2013nza,Drischler:2020hwi,Hagen:2015yea,Chen:2010qx,Steiner:2011ft,
Gandolfi:2013baa,Roca-Maza:2015eza}. We underscore that the models used in this letter represent 
a particular class of relativistic EDFs. 

We note that theoretical interpretations 
of elastic nucleon-nucleus scattering cross sections together with quasielastic reactions to isobaric analog
states obtained limits on $L$ that are consistent with our findings\,\cite{Danielewicz:2016bgb}. 
The PREX-II result is also considerably larger---and in many cases incompatible---with 
experimental determinations of \Rskin by methods that are highly model 
dependent\,\cite{Trzcinska:2001sy,Zenihiro:2010zz,Tarbert:2013jze,Thiel:2019tkm}. 
A notable exception is the dispersive optical model analysis of the Washington 
University group that reported a neutron skin thickness of 
\Rskin$\!=\!(0.25\pm0.05)\,{\rm fm}$\,\cite{Atkinson:2019bwd}; a revised lower value of
\Rskin$\!=\!(0.18\pm0.07)\,{\rm fm}$---still consistent with\,\cite{Atkinson:2019bwd}---was reported 
shortly thereafter in Ref.\,\cite{Pruitt:2020zbf}.

\begin{figure}[ht]
 \centering
 \includegraphics[width=7cm,height=8.5cm]{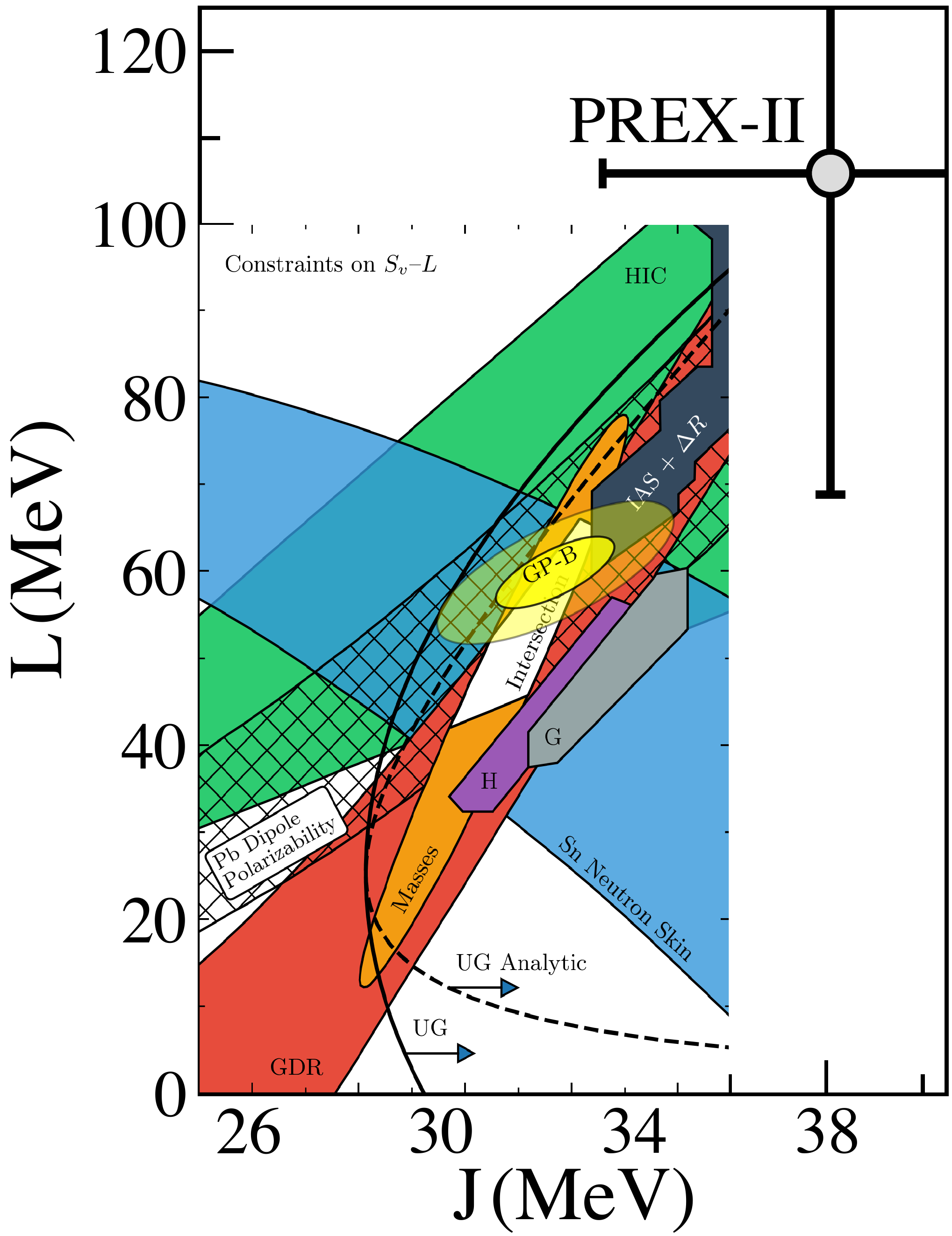}
 \caption{(Color online). Constraints on the $J$--$L$ correlation obtained from a variety of experimental and 
 theoretical approaches. The figure was adapted from Refs.\,\cite{Lattimer:2012nd,Drischler:2020hwi} and 
 noticeably displays the tension with the recent PREX-II result.}
\label{Fig2}
\end{figure}

To further underscore the tension between PREX-II and our current understanding of the EOS, 
we display in Fig.\,\ref{Fig2} a summary of simultaneous constraints on 
both $J$ and $L$ as reported in Refs.\,\cite{Lattimer:2012nd,Drischler:2020hwi}. We have
adapted Figure\,2 from Ref.\,\cite{Drischler:2020hwi} by including the PREX-II limits on both 
$J$ and $L$ derived in Eq.(\ref{JandL}). Note that with the exception of the analysis of 
Ref.\,\cite{Chen:2010qx}, all other approaches suggest a positive correlation between $L$ and $J$. 
In the context of density functional theory, such a positive correlation is easy to understand. 
Using Eq.(\ref{EandS}) at $\widetilde\rhozero$ yields
\begin{equation}
 S(\widetilde\rhozero) = J - \frac{L}{9} \rightarrow 
 J \approx \left(26\,{\rm MeV} + \frac{L}{9}\right).
 \label{Jtilde}
\end{equation}
The value of $S(\widetilde\rhozero)\!\approx\!26\,{\rm MeV}$\,\cite{Horowitz:2000xj} follows
because the symmetry energy at $\widetilde\rhozero$ is tightly constrained by the 
binding energy of heavy nuclei. The PREX-II inferred value for $L$ yields a corresponding value 
of $J\!=\!(37.7\pm 4.1)\,{\rm MeV}$, that is entirely consistent with the limit obtained in 
Eq.(\ref{JandL}). 
Although consistent at the $2\sigma$ level,
the ``Intersection" region in Fig.\,\ref{Fig2} obtained from a variety of experimental and theoretical 
approaches lies outside the $1\sigma$ PREX-II limits.

\begin{figure}[ht]
 \centering
 \includegraphics[width=0.35\textwidth]{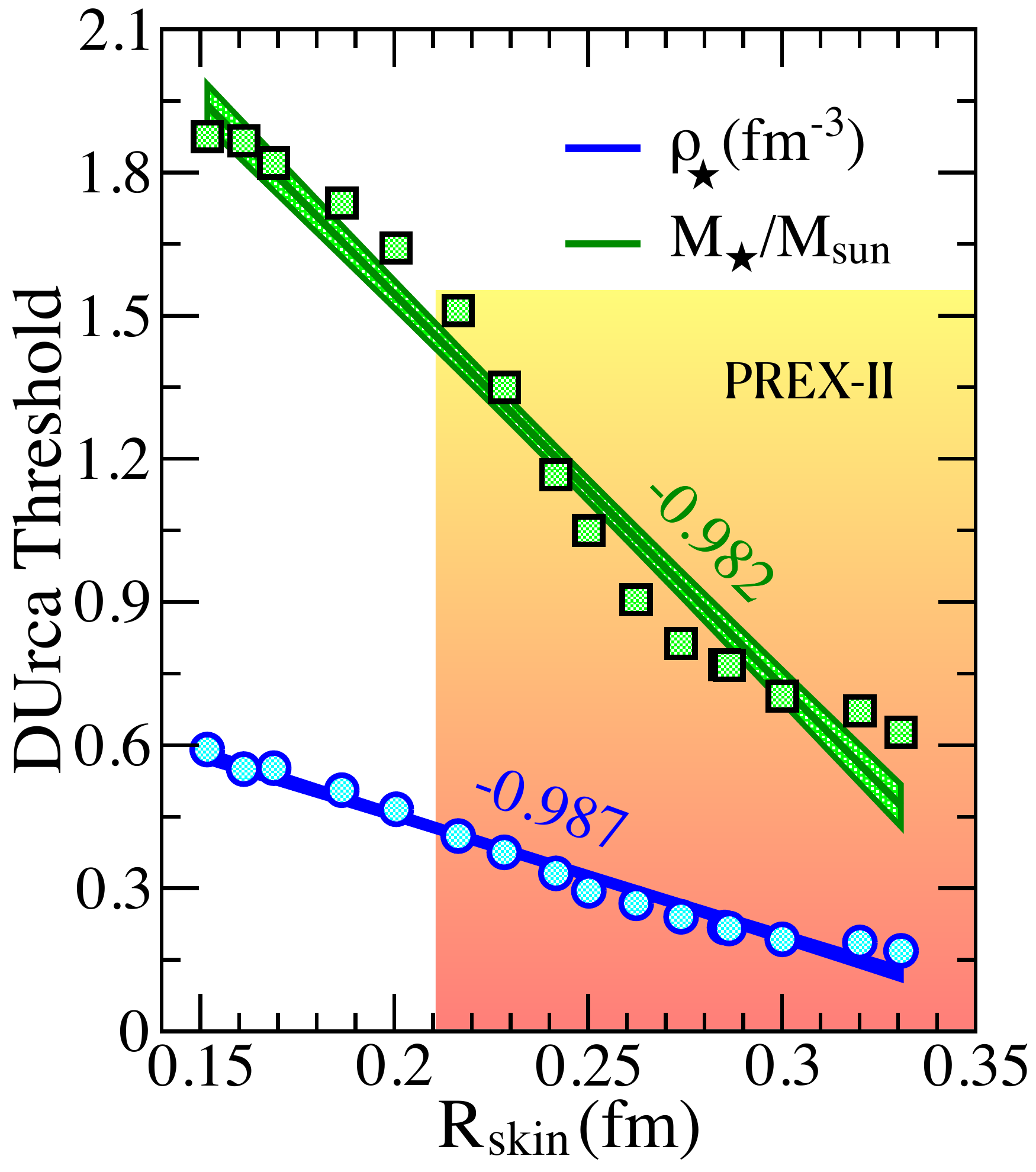}
 \caption{(Color online). Direct Urca thresholds for the onset of enhanced cooling in neutron stars. The threshold 
 density is depicted by the lower blue line and the corresponding stellar mass for such a central density with the 
 upper green line. The shaded area represents PREX-II $1\sigma$ confidence region. For each of these two 
 quantities, the best-fit line is displayed together with their associated correlation coefficients.}
\label{Fig3}
\end{figure}

Next, we explore the impact of PREX-II on a few neutron-star observables. We start by displaying in Fig.\,\ref{Fig3} 
the minimum central density and associated neutron star mass required for the onset of the direct Urca process. 
Neutron stars are born very hot ($T\!\simeq\!10^{11}K\!\simeq 10\,{\rm MeV}$) and then cool rapidly via neutrino 
emission through the direct Urca process that involves neutron beta decay followed by electron capture:
\begin{subequations}
\begin{align}
 & n \rightarrow p + e^{-} + \bar{\nu}_{e}, \\
 & p + e^{-} \rightarrow n + \nu_{e}.
\end{align} 
\label{DUrca}
\end{subequations}
After this rapid cooling phase is completed, neutrino emission proceeds in the standard cooling scenario through 
the modified Urca process---a process that may be millions of times slower as it  requires the presence of a 
bystander nucleon to conserve momentum at the Fermi surface\cite{Page:2004fy}. The transition into the much 
slower modified Urca process is solely based on the expectation that the proton fraction in the stellar core is too 
low to conserve momentum at the Fermi surface. However, given that the proton fraction is controlled by the poorly 
known density dependence of the symmetry energy\,\cite{Horowitz:2002mb}, the minimal cooling scenario may 
need to be revisited. In particular, a stiff symmetry energy---as suggested by PREX-II---favors large proton fractions 
that may trigger the onset of the direct Urca process at lower central densities. This analysis is particularly timely 
given that x-ray observations suggest that some neutron stars may  require some form of enhanced cooling. 
Indeed, the detected x-ray spectrum of the neutron star in the low-mass x-ray binary MXB 1659-29 strongly suggests
the need for a fast neutrino-cooling process\,\cite{Brown:2017gxd}. For a comprehensive report that explores the 
interplay between the direct Urca process and nucleon superfluidity in transiently accreting neutron stars, see 
Ref.\,\cite{Potekhin:2019eya}. The shaded area in Fig.\,\ref{Fig3} displays the region constrained by PREX-II. 
In particular, the $1\sigma$ lower limit of $R_{\rm skin}\!=\!0.212\,{\rm fm}$ suggests a threshold mass for the 
onset of direct Urca cooling of $M_{\star}\!\approx\!1.45\,M_{\odot}$ and a corresponding central density of 
$\rho_{\star}\!\approx\!0.42\,{\rm fm}^{-3}$. However, if instead one adopts the larger PREX-II central value of 
$R_{\rm skin}\!=\!0.283\,{\rm fm}$, then one obtains the considerably lower threshold values of 
$M_{\star}\!\approx\!0.85\,M_{\odot}$ and $\rho_{\star}\!\approx\!0.24\,{\rm fm}^{-3}$, or a threshold density 
just slightly higher than saturation density. Although some stars are likely to require enhanced cooling, observations 
of many isolated neutron stars are consistent with the much slower modified URCA process\,\cite{Page:2009fu}. 
This may be because the direct URCA neutrino emissivity is reduced by nucleon pairing.        

\begin{figure}[ht]
 \centering
 \includegraphics[width=0.4\textwidth]{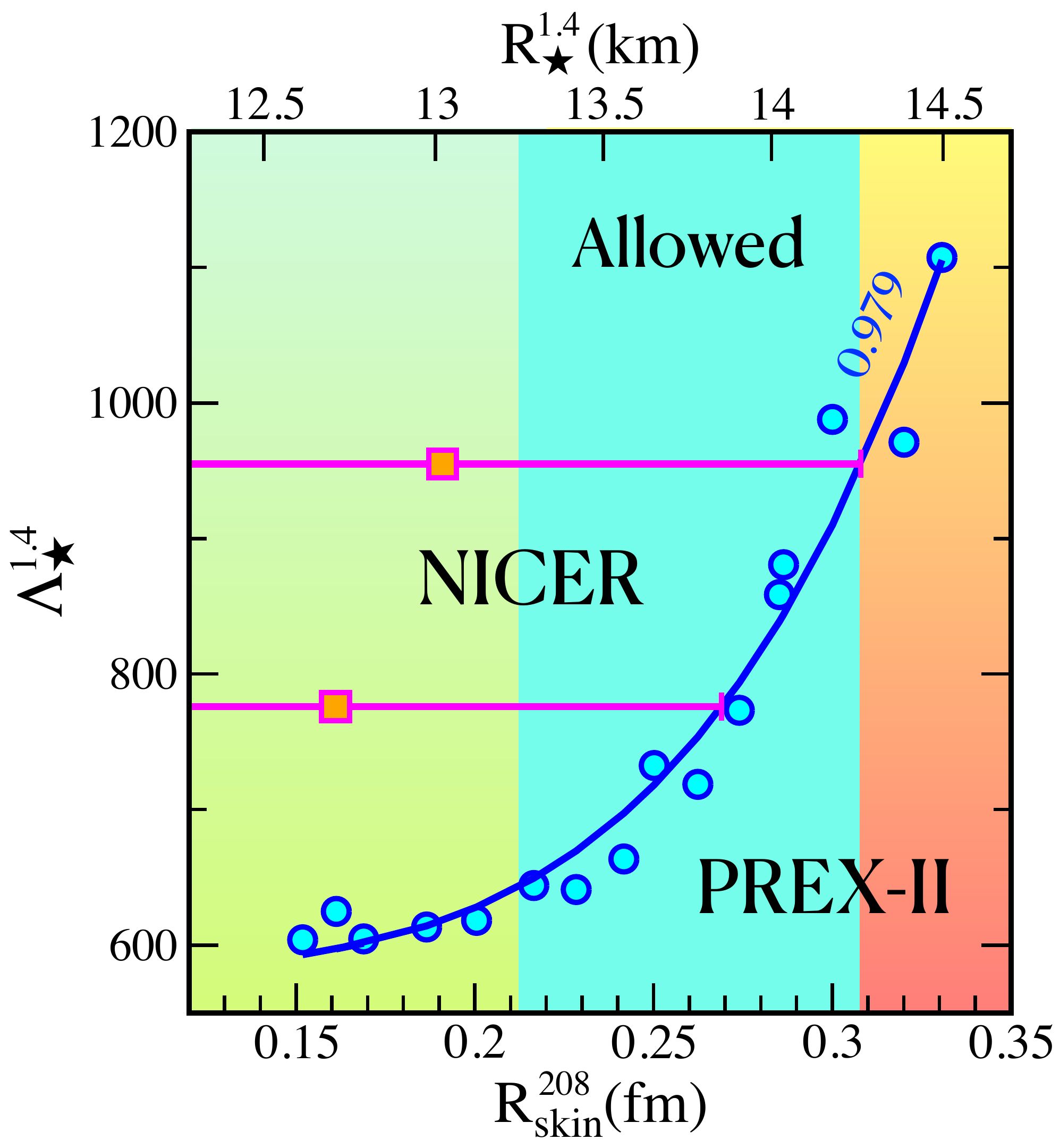}
 \caption{(Color online). Showcase of neutron star observables as a function of \Rskin \ as predicted
 by the set of energy density functionals considered in this work. The tidal deformability \Lns of a 
 1.4 \msun \ neutron star 
 is computed for each model, 
 and displayed with blue dots and connected by 
 a best fit power law that scales as the $4.8\!\approx\!5$ power of \Rns.  The combined PREX-II result 
 together with NICER constraints on the stellar radius is depicted by the small (blue) window of models 
 allowed.}
\label{Fig4}
\end{figure}

We close the section by displaying in Fig.\,\ref{Fig4} the dimensionless tidal deformability of a 1.4\,$M_{\odot}$ 
neutron star as a function of both the stellar radius $R_{\star}^{1.4}$ and $R_{\rm skin}$. Although not shown, 
for the set of density functionals used in this work a very strong correlation (of about 0.98) is obtained between 
$R_{\star}^{1.4}$ and $R_{\rm skin}$. However, because the central density of a 1.4\,$M_{\odot}$ neutron star 
may reach densities as high as 2-to-3 times saturation density, the robustness of such a correlation should be 
examined in the context of alternative theoretical descriptions. Moreover, a precise knowledge of the EOS of the 
crust is needed to minimize possible systematic uncertainties\,\cite{Piekarewicz:2014lba}. As in Fig.\,\ref{Fig3}, 
the $1\sigma$ confidence region is indicated by the shaded area in the figure. Also shown are NICER constrains
on the radius of PSR J0030+0451\,\cite{Riley:2019yda,Miller:2019cac}, that are depicted by the two horizontal 
error bars and which suggest an upper limit of $R_{\star}^{1.4}\!\le\!14.26\,{\rm km}$. Invoking the strong 
$R_{\star}^{1.4}$\,--\,$R_{\rm skin}$ correlation observed in our models, one obtains an upper limit on the neutron 
skin thickness of $R_{\rm skin}\!\lesssim\!0.31\,{\rm fm}$ and a lower limit on the stellar radius of 
$R_{\star}^{1.4}\!\gtrsim\!13.25\,{\rm km}$. The region that satisfies both PREX-II and NICER constraints is 
indicated by the narrow (blue) rectangle in Fig.\,\ref{Fig4}, which excludes a significant number of models. In 
turn, given that the tidal deformability approximately scales with the fifth power of the stellar 
radius\,\cite{Fattoyev:2017jql}, one can also set limits on the tidal deformability of a $1.4\,M_{\odot}$ neutron star. 
Combining the constraints from NICER on \Rns and PREX-II on \Rskin one obtains:
\begin{subequations}
\begin{align}
    0.21 \lesssim & \; R _{\rm skin}({\rm fm}) \!\lesssim 0.31 \\
    13.25 \lesssim & \; R _{\star}^{1.4}({\rm km}) \!\lesssim 14.26 \\
    642 \lesssim & \; \Lambda_{\star}^{1.4} \!\lesssim 955.
\end{align} 
\label{NStars}
\end{subequations}
The allowed region for the tidal deformability falls comfortably within the $\Lambda_{\star}^{1.4}\!\lesssim\!800$
limit reported in the GW170817 discovery paper\,\cite{Abbott:PRL2017}. Yet, the revised limit of 
$\Lambda_{1.4}\!=\!190^{+390}_{-120}\!\lesssim\!580$\,\cite{Abbott:2018exr} 
presents a more serious challenge. To confirm whether this tension is real, it will require a multi-prong approach
involving a more precise determination of \Rskin, additional NICER observations, and more multi-messenger 
detections of neutron star mergers. The prospect of a more precise electroweak determination of \Rskin\, is 
challenging as it may require the full operation of the future Mainz Energy-recovery Superconducting 
Accelerator (MESA) which is foreseen to start until 2023\,\cite{Becker:2018ggl}. Future determinations of stellar 
radii by NICER for neutron stars with known masses, such as J0437-4715\,\cite{Reardon:2015kba}, could be 
made at a $\pm\,3\%$ level, or to better than $\pm\,0.5\,{\rm km}$.  NICER is also collecting pulse profile 
modeling data for the highest mass pulsar (PSR J0740+6620) ever measured\,\cite{Cromartie:2019kug}. Finally, 
the LIGO-Virgo-KAGRA collaborations are preparing for the fourth observing run at a higher detector sensitivity. 
Although KAGRA will join LIGO and Virgo promising much better sky localization, COVID-related delays have 
pushed the fourth observing run until June 2022. 

In summary, PREX-II has confirmed with improved precision the original PREX suggestion that the EOS 
at the typical densities found in atomic nuclei is stiff. This result challenges our present understanding of 
the density dependence of symmetry energy extracted from various experimental and theoretical 
analyses\,\cite{Thiel:2019tkm}.
By assessing the impact of PREX-II at higher densities, we were able to provide
limits on both the radius and deformability of a 1.4\,$M_{\odot}$ neutron star. Given that
our analysis of the tidal deformability reveals some tension with the revised limit of 
$\Lambda_{1.4}\!\lesssim\!580$\,\cite{Abbott:2018exr}, we eagerly await the next generation of terrestrial 
experiments and astronomical observations to verify whether the tension remains. If so, the softening of the
EOS at intermediate densities, together with the subsequent stiffening at high densities required to support 
massive neutron stars, may be indicative of a phase transition in the stellar core\,\cite{Fattoyev:2017jql}.

\medskip
\begin{acknowledgments}\vspace{-10pt}
 This material is based upon work supported by the U.S. Department of Energy Office of Science, Office of 
 Nuclear Physics under Awards DE-FG02-87ER40365 (Indiana University), Number DE-FG02-92ER40750 
 (Florida State University), and Number DE-SC0008808 (NUCLEI SciDAC Collaboration).
\end{acknowledgments} 

\bibliography{PREX-II.bbl}
\end{document}